\documentstyle[preprint,aps,tighten,epsf]{revtex}

\begin{document}
\draft
\title{Chaos in one-dimensional lattices under intense laser fields} 

\author{M. A. M. de Aguiar$^a$, H. S. Brandi$^b$, Belita Koiller$^b$, and
Eduardo R. Mucciolo$^c$}
\address{$^a$ Instituto de F\'\i sica, Universidade Estadual de Campinas,}
\address{Cx. P. 6165, 13083-970, S\~ao Paulo, Brazil}

\address{$^b$ Instituto de F\'\i sica, Universidade Federal do Rio de 
Janeiro,}
\address{Cx. P. 68.528, 21945-970, Rio de Janeiro, Brazil}
\address{$^c$ Departamento de F\'\i sica, Pontif\'\i cia 
Universidade Cat\'olica do Rio de Janeiro,}
\address{Cx. P. 38071, 22452-970, Rio de Janeiro, Brazil}

\date{\today}

\maketitle

\begin{abstract}
A model is investigated where a monochromatic, spatially homogeneous
laser field interacts with an electron in a one-dimensional periodic
lattice. The classical Hamiltonian is presented and the technique of
stroboscopic maps is used to study the dynamical behavior of the
model. The electron motion is found to be completely regular only for
small field amplitudes, developing a larger chaotic region as the
amplitude increases. The quantum counterpart of the classical
Hamiltonian is derived. Exact numerical diagonalizations show the
existence of universal, random-matrix fluctuations in the electronic
energy bands dressed by the laser field. A detailed analysis of the
classical phase space is compatible with the statistical spectral
analysis of the quantum model. The application of this model to
describe transport and optical absorption in semiconductor
superlattices submitted to intense infrared laser radiation is
proposed.

\end{abstract}
\narrowtext

%%%%%%%%%%%%%%%%%%%%%%%%%%%%%%%%%%%%%%%%%%%%%%%%%%%%%%%%%%%%%%%%%%%%%%%%%
\section{Introduction}
\label{sec:1}

The search for systems whose dynamical behavior can be fine tuned by
one or more external parameters has been an important activity in the
field of chaos in the last decades \cite{reviewchaos}. The kicked
rotor \cite{casati79} has emerged as the paradigm model of a
periodically driven system whose motion depends strongly on the
strength of the external perturbation. The study of this particular
model allowed us to understand phenomena such as the dynamical
localization of atoms interacting with microwave radiation in an
optical trap \cite{Natrapped}.

For time-independent, Hamiltonian systems, one has also extensively
searched for situations where the dynamics is a sensitive function of
a {\it single} parameter. Common situations occur when two integrable
potentials are coupled to form a nonintegrable system. One case of
particular interest, both theoretically and experimentally, is the
hydrogen atom in the presence of a uniform magnetic field
\cite{hatom}, since the limiting cases of zero magnetic field and zero
Coulomb force are exactly solvable.

In this work we study a model describing the motion of an electron in
a one-dimensional periodic potential in the presence of an external
monochromatic laser field. The most accessible experimental
realization of this model occurs in semiconductor superlattices
(multiple quantum-well structures) irradiated with intense infrared
laser pulses.

We have considered a one-dimensional time-periodic effective classical
Hamiltonian describing the interaction of an intense laser field with
an electron in a 1-D periodic lattice. We show that the effective
Hamiltonian is {\it integrable} in the limiting case of zero field
strength. As the field strength becomes larger, the behavior of the
classical orbits become increasingly chaotic. For strong fields the
Hamiltonian is invariant under a scaling transformation involving the
electron momentum, the average energy of the Hamiltonian, and the
field strength. Therefore, to understand the whole range of laser
intensities it is necessary only to study the weak and intermediate
field regimes.

In order to propose a physical realization of this model situation we
have studied the quantum counterpart of the classical
Hamiltonian. Through numerical simulations, we found that typical
signatures of quantum chaos, such as level repulsion and spectral
rigidity, are present for intermediate and large values of field
strengths. Based on the energy level statistics observed, it was
possible to identify a weak (Poisson) and a strong (Wigner-Dyson)
level correlation regime as the laser intensity increases. The
statistical spectral analysis approach, which provides strong evidence
of chaotic and regular behaviors, can be completely understood in the
light of the classical model.

This paper is organized as follows. In Sec.~\ref{sec:2} we present the
classical model and study the dynamics of the system as a function of
the laser field strength using the technique of stroboscopic maps. In
Sec.~\ref{sec:3} we derive the quantum model and compare the
classical results. A discussion on possible experimental observations
is given in Sec.~\ref{sec:4}. Finally, Sec.~\ref{sec:5} is dedicated
to the conclusions.

%%%%%%%%%%%%%%%%%%%%%%%%%%%%%%%%%%%%%%%%%%%%%%%%%%%%%%%%%%%%%%%%%%%%%%%
\section{The Classical Model}
\label{sec:2}

Consider the Hamiltonian describing an electron in a 1-D periodic
potential under the action of an applied homogeneous monochromatic
laser field. The laser field can be represented by a time-dependent
vector potential (in the dipole approximation) minimally coupled to
the electron. In this limit, the laser field is not affected by the
electron motion and the vector potential acts as a time-periodic
external force. The lattice potential is periodic and can be written
in terms of a sum over reciprocal vectors $G_\ell = 2\pi \ell /d$,
with $d$ denoting the lattice constant and $\ell =0,\pm 1,\pm
2,\ldots$. Explicitly,
\begin{equation}
H(p,q) = \frac{1}{2m} \left[ p - \frac{e}{c} A_0 \cos (\omega t)
\right]^2 - 2 \sum_{\ell=1}^{\infty} \sigma_\ell \cos \left(\ell G_1 q
\right) .
\end{equation}
The term proportional to $A_0^2$ in $H$ yields a pure time-dependent
factor that does not affect the equations of motion of the electron.
Therefore, this term can be eliminated and the classical motion can be
obtained from the effective Hamiltonian
\begin{equation}
H_{\scriptsize{\mbox{eff}}}(p,q) = \frac{p^2}{2m} - 2
\sum_{\ell=1}^{\infty} \sigma_\ell \cos \left(\ell G_1 q \right) +
\frac{e A_0 p}{m c} \cos (\omega t).
\label{eq:effH}
\end{equation}
In order to compare the results of the classical calculations with the
quantum analysis shown bellow, we first introduce suitable energy and
frequency scales into the problem: $\epsilon \equiv \hbar^2 G_1^2/2m$
and $\hbar \omega = \epsilon \omega_0$, such that the dimensionless
lattice potential strength becomes $\sigma_\ell^0 \equiv
\sigma_\ell/\epsilon$. This naturally leads to a new pair of
canonically conjugate variables
\begin{displaymath}
P = \frac{\hbar}{\sqrt{m \epsilon}} p
\end{displaymath}
and
\begin{displaymath}
Q = \frac{\sqrt{m \epsilon}}{\hbar} q
\end{displaymath}
and a new time scale $\tau \equiv \epsilon t /\hbar$. The resulting
equations of motion for $P(\tau)$ and $Q(\tau)$ can be derived from
the rescaled Hamiltonian
\begin{equation}
{\cal H}(P,Q) = \frac{(P/\hbar)^2}{2} - 2 \sum_{\ell=1}^{\infty}
\sigma_\ell^0 \cos \left(\ell \sqrt{2} Q \right) + \frac{x \omega_0
(P/\hbar)}{\sqrt{2}} \cos (\omega_0 \tau),
\label{hamilclass}
\end{equation}
where $x \equiv e A_0 G_1/\omega mc$. The total energy associated to
the effective Hamiltonian in Eq.~(\ref{eq:effH}) is related to the
rescaled (dimensionless) energy ${\cal E}(\tau)$ by
$E_{\scriptsize{\mbox{eff}}} (t) = \epsilon \; {\cal E}(\epsilon t
/\hbar)$. The dimensionless parameter $x$ characterizes the laser
intensity. For numerical estimates, it can be expressed in the more
appropriate form
\begin{equation}
x = \left( \frac{\lambda}{d} \right) \sqrt{\frac{n_r I}{ I_C}},
\label{eqx}
\end{equation}
with $I = n_r \omega^2 A_0^2/8 \pi c$ and $n_r$ denoting the
refraction index of the medium. The quantity $I_C = m^2 \omega^2
c^3/8\pi e^2$ is the Compton intensity, i.e., $I_C = 1.37 \times
10^{18} \times \lambda^{-2}\ [\mbox{W}/\mbox{cm}^2]$, with the laser
wavelength $\lambda$ given in $\mu$m.

Since the potential is a periodic function of $Q$, $Q$ and $P$ are
{\em angle} and {\it action} variables, respectively, with $P$ having
units of $\hbar$. Thus, hereafter, we shall drop the $\hbar$ in
Eq.~(\ref{hamilclass}) and understand that $P$ is measured in units of
$\hbar$. We shall also restrict $Q$ to the primitive cell interval
$-\pi /\sqrt{2}$ to $\pi /\sqrt{2}$.

In order to carry out our analysis, we will consider only the first
three reciprocal components of the periodic lattice potential in
Eq.~(\ref{hamilclass}). There is nothing special about this choice and
it just represents a possible truncation of the well-known
Kronig-Pennig potential. The number of components is taken to be
relatively small in order to facilitate the numerical simulations; at
the same time, it is also sufficiently large to allow for the
existence of a nontrivial classical dynamics. For similar reasons, in
all numerical calculations (classical and quantum) in this work we
shall adopt $\sigma_\ell^0 = \sigma^0 = 0.25$ (for $-3\le \ell \le 3$)
and $\omega_0 = 0.3$.

\begin{figure}
\setlength{\unitlength}{1mm}
\begin{picture}(160,140)(0,0)
\put(0,10){\epsfxsize=60mm\epsfbox{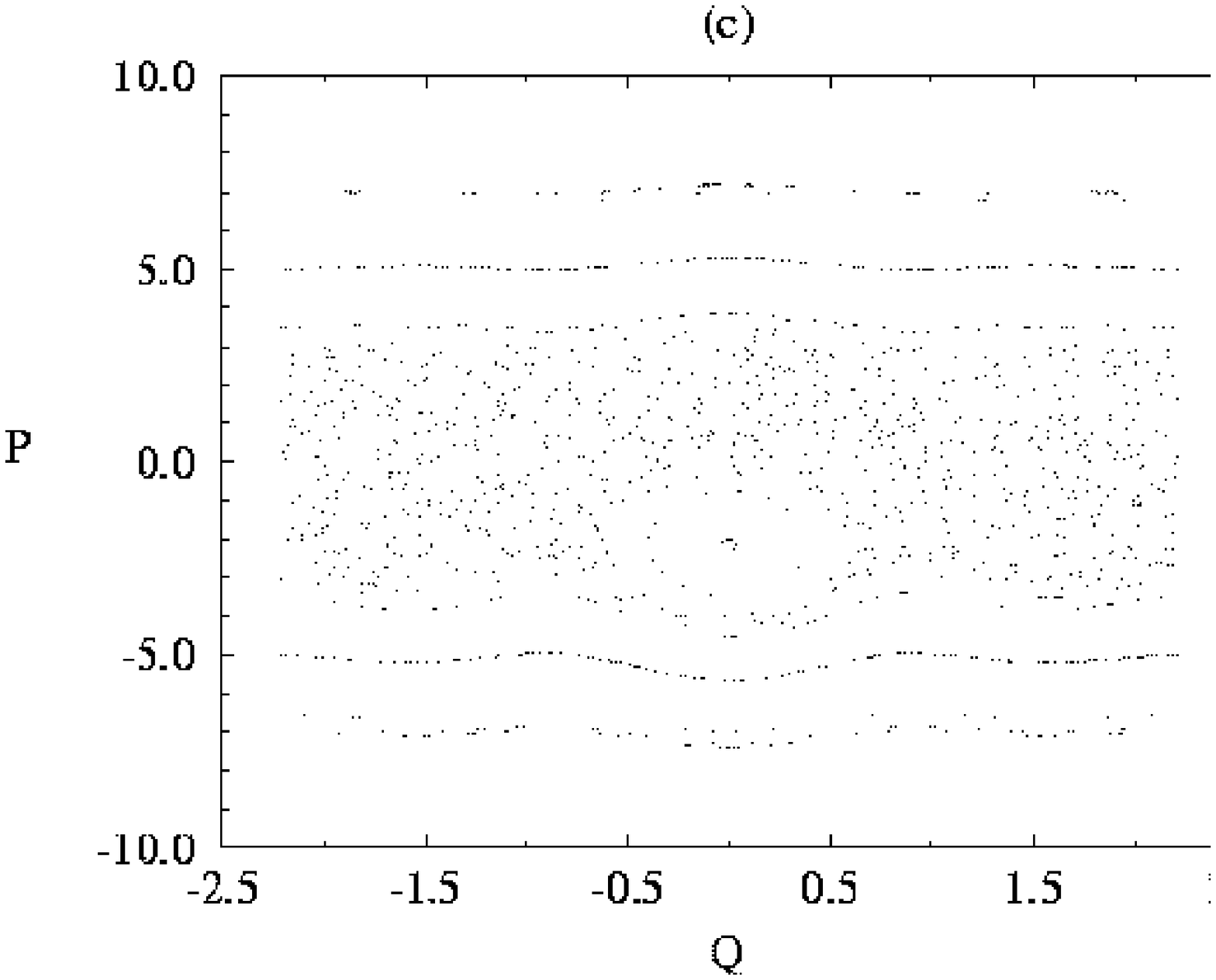}}
\put(80,10){\epsfxsize=60mm\epsfbox{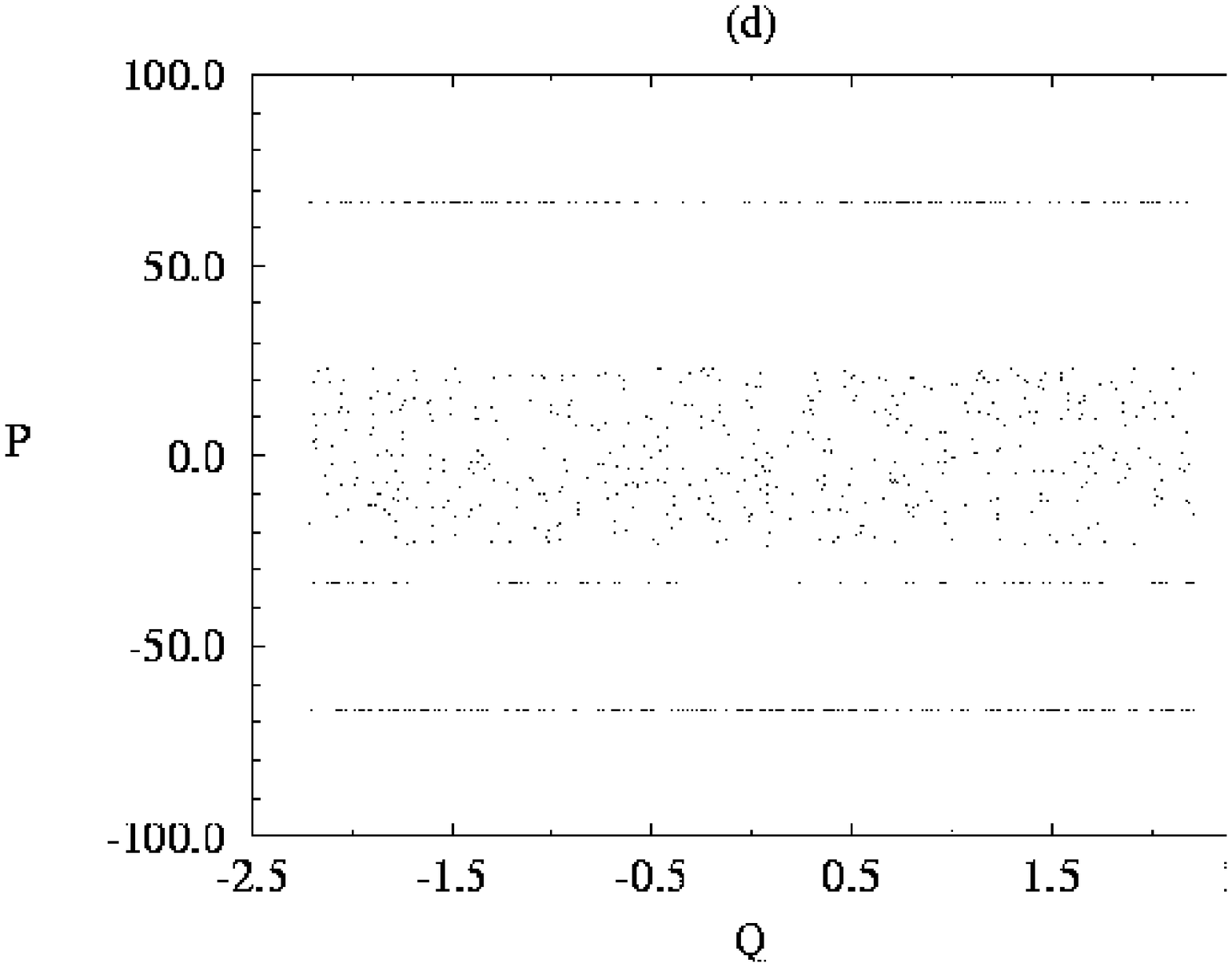}}
\put(0,70){\epsfxsize=60mm\epsfbox{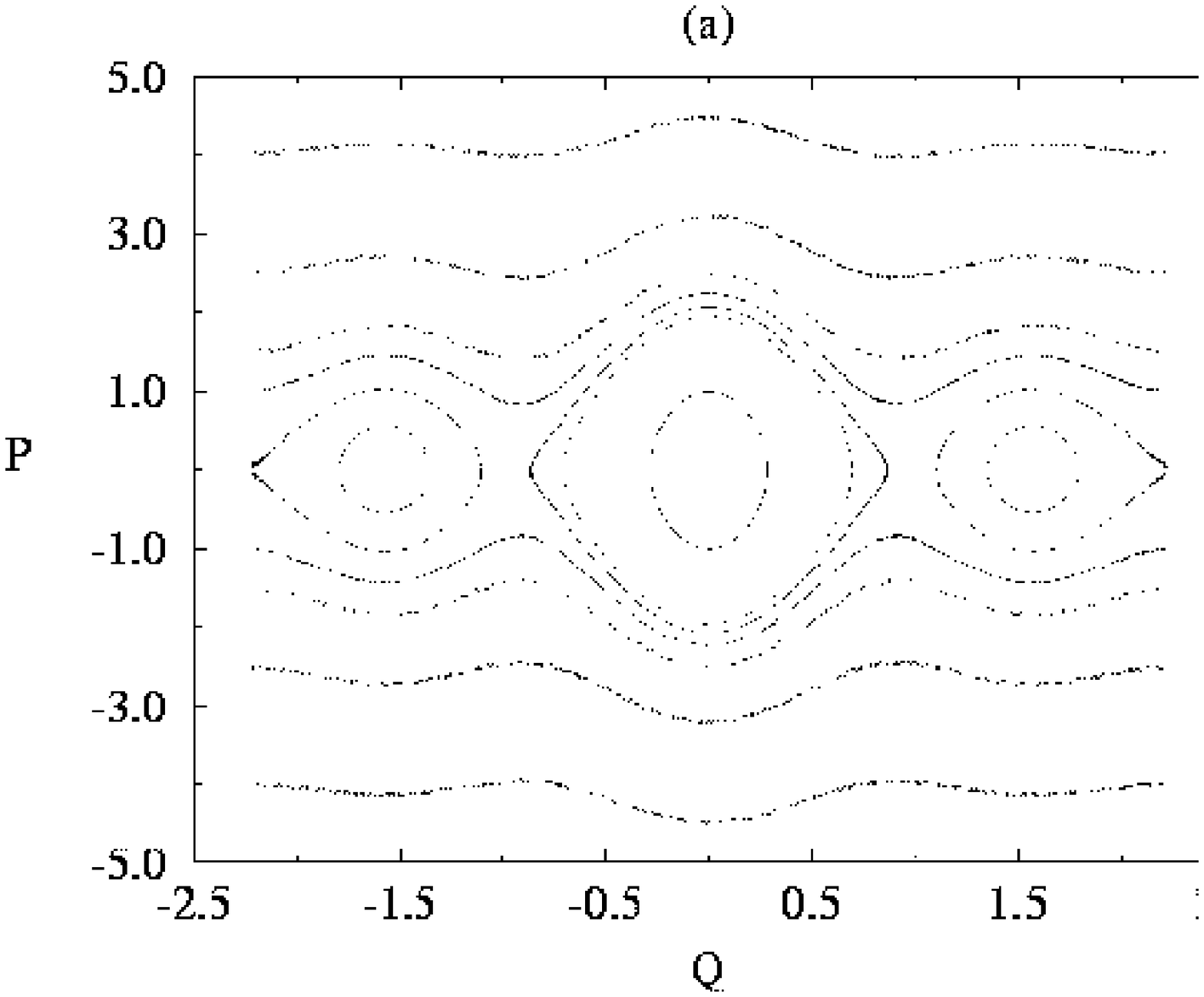}}
\put(80,70){\epsfxsize=60mm\epsfbox{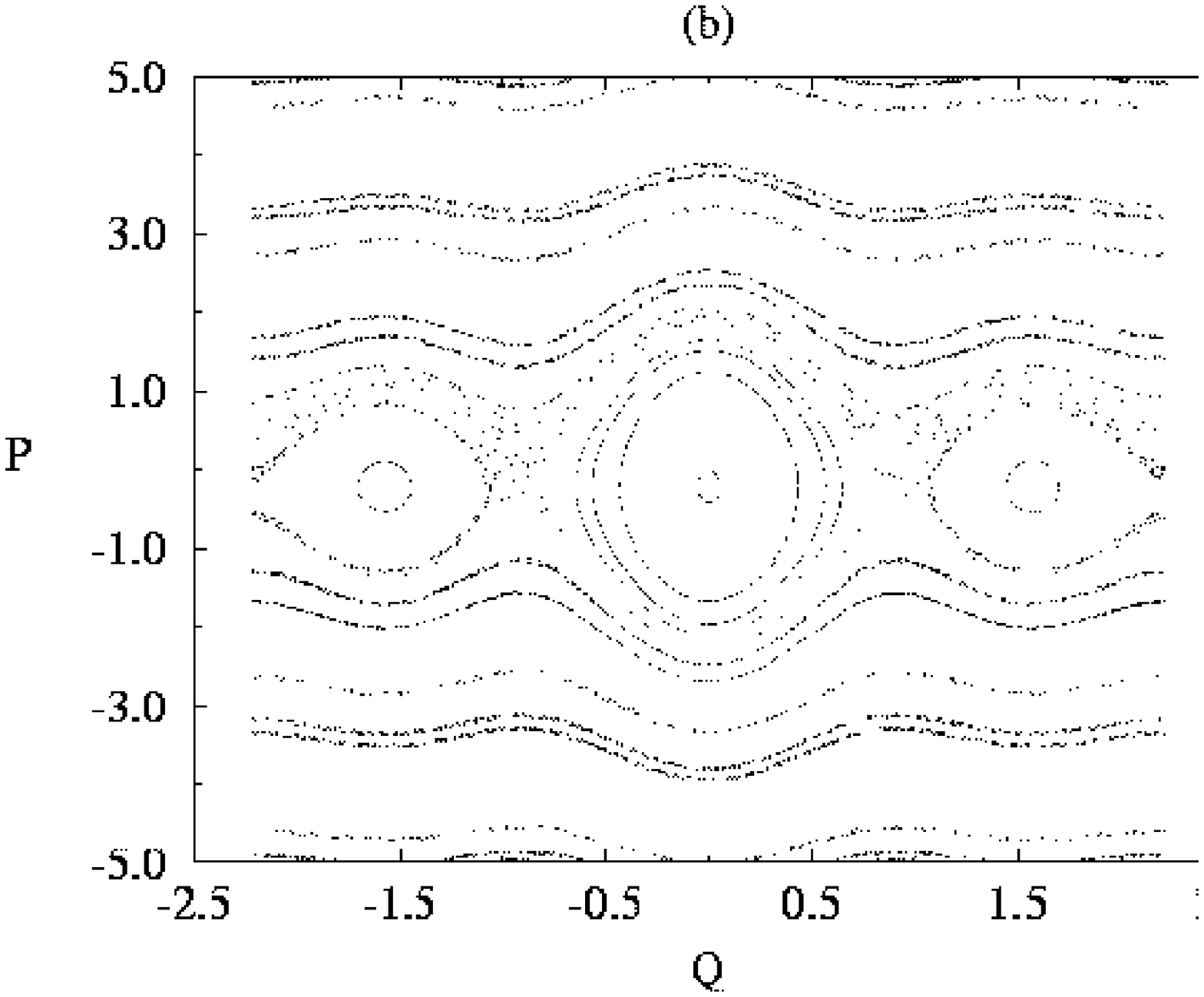}}
\end{picture}
\caption{Stroboscopic maps of the Hamiltonian in
\protect{Eq.~(\ref{hamilclass})} for (a) $x=0$, (b) $x=1$, (c) $x=10$, and
(d) $x=100$. See the text for a discussion.}
\label{fig1}
\end{figure}

\begin{figure}
\setlength{\unitlength}{1mm}
\begin{picture}(120,80)(0,0)
\put(20,10){\epsfxsize=80mm\epsfbox{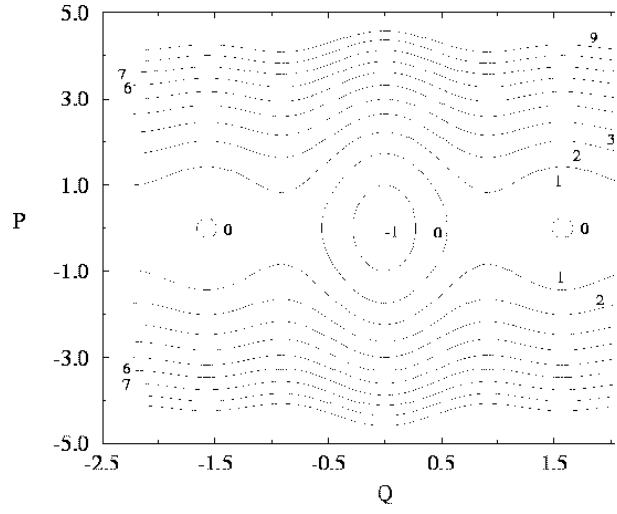}}
\end{picture}
\caption{Contour plots of the Hamiltonian in
\protect{Eq.~(\ref{hamilclass})} for $x=0$.}
\label{fig2}
\end{figure}

Figure~\ref{fig1} displays {\it stroboscopic maps} for the classical
system defined by Eq.~(\ref{hamilclass}) at different laser
intensities. These maps are generated by plotting the coordinates $Q$
and $P$ at discrete times $\tau_n = 2 \pi n/\omega_0$,
$n=0,1,2,\ldots$, producing a sequence of points $(Q_n$,$P_n)$. Each
initial condition $(Q_0$,$P_0)$ gives rise to a different sequence. If
these points fall on a one-dimensional curve, the trajectory lies on a
cylinder in the extended ($Q,P,\tau$) phase space and is said to be
regular. If, on the other hand, the points cover a two-dimensional
area of the $QP$ plane, the trajectory is said to be irregular or
chaotic. Figure~\ref{fig2} shows contour levels of the Hamiltonian
for $x=0$.

Let us now discuss the classical motion arising from
Eq.~(\ref{hamilclass}) in terms of stroboscopic maps for different
laser intensities. For $x=0$, Figure~\ref{fig1}(a), the map is
integrable and all points lie on 1-D curves. The wells centered at
$Q=P=0$ and $P=0$, $Q \simeq \pm 1.5$, trap low-energy electrons,
whereas those with higher energies, above or bellow these trapping
{\it islands}, move along the lattice from left to right (positive
$P$) or from right to left (negative $P$). These trajectories are
clearly related to the conduction properties of the material. In
Figure~\ref{fig2} the numbers close to the curves indicate the value
of ${\cal E}$. For $x \neq 0$, only average energies can be defined:
above the islands, they are roughly equal to the kinetic term alone,
since the other terms have an oscillatory behavior.

\begin{figure}
\setlength{\unitlength}{1mm}
\begin{picture}(100,100)(0,0)
\put(20,10){\epsfxsize=80mm\epsfbox{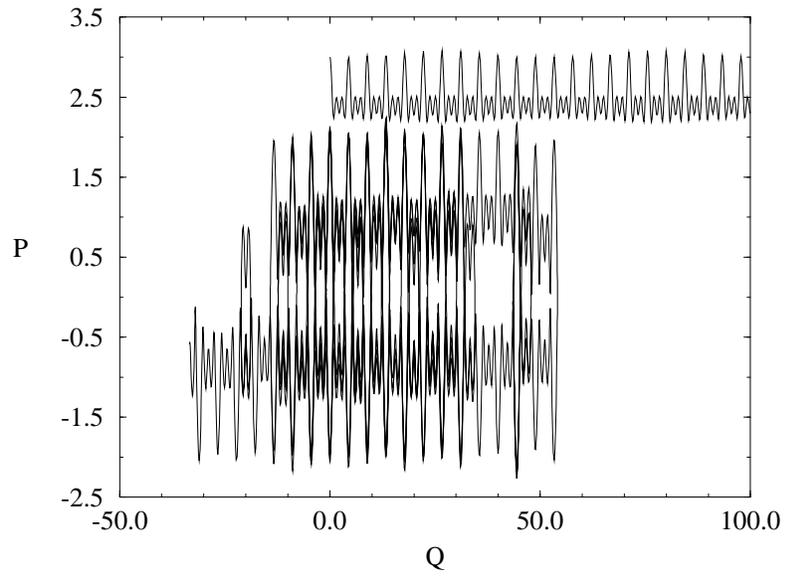}}
\end{picture}
\caption{Classical electron trajectories for $x=1$. The upper curve is
a regular trajectory for the initial conditions $Q=0$, $P=3$ and
propagation time = 50. The lower curve is a chaotic trajectory for the
initial conditions $Q=0$, $P=1.9$ and propagation time = 500.}
\label{fig3}
\end{figure}

For slightly larger intensities, a small amount of chaos begins to
permeate the region where the separatrices involving the islands used
to be. This effect is clearly visible in Figure~\ref{fig1}(b), where
$x=1$. In the new chaotic zone, the electromagnetic field {\it shakes}
the electron, forcing it to move in opposite directions
intermittently, inhibiting charge transport through the
crystal. Figure~\ref{fig3} shows typical electron trajectories for
chaotic (lower curve) and regular (upper curve) regimes when $x=1$.
These trajectories are related to different initial conditions. Note
that the chaotic trajectory is restricted to a narrower range of
$Q$-values in comparison to the regular trajectory, which extends over
a much wider $Q$ region.

For $x=10$, Fig~.(\ref{fig1}c), a rather large chaotic zone develops
at $|P| \lesssim 3.5$. Figure~\ref{fig1}(d) shows that for $x=100$,
the region close to $|P|=0$ becomes dominated by chaotic
trajectories. Notice that the maps for $x=10$ and $x=100$ are almost
identical at large $|P|$. This is due to the existence of a {\it
quasi} scaling law in the classical motion: For $x\gg 1$, one can
discard the periodic potential and change $P \rightarrow \alpha P$ and
$x \rightarrow \alpha x$, rescaling ${\cal E} \rightarrow
\alpha^2 {\cal E}$. As a consequence, it is possible to infer that for
very strong field intensities the regime will be predominantly
chaotic, even at (relatively) low energies.

From the figures and the equations of motion, for any given value of
$x$, it is easy to see that $P$ tends to a constant, provided its
initial value is sufficiently large. Moreover, at $x=0$ the average
energy also becomes equal to a constant since the average of the
coupling term becomes negligible. The small $P$ region, on the other
hand, is largely affected by the electromagnetic field and the lattice
potential if $x$ is not too small.

%%%%%%%%%%%%%%%%%%%%%%%%%%%%%%%%%%%%%%%%%%%%%%%%%%%%%%%%%%%%%%%%%%%%%%%%
\section{The Quantum Model}
\label{sec:3}

The quantum Hamiltonian corresponding to the classical model of the
previous section can be written in the form
\begin{equation}
\label{eq:tothamilt}H = H_k + H_{\scriptsize{\mbox{int}}},
\end{equation}
where
\begin{eqnarray}
H_k & = & \frac{p^2}{2m} - 2 \sum_{\ell=1}^{3} \sigma_\ell \cos
\left(\ell G_1 q \right) \nonumber \\ & = & -\frac{\hbar^2}{2m} \left(
{d\over {dq}} + ik \right)^2 - 2 \sum_{\ell=1}^{3} \sigma_\ell \cos
\left(\ell G_1 q
\right)
\label{eq:elechamilt}
\end{eqnarray} 
and
\begin{equation}
\label{eq:interhamilt}H_{\scriptsize{\mbox{int}}} = \frac e{mc}\,A(t)p
+ \frac{e^2}{2mc^2} A(t)^2,
\end{equation}
with
\begin{equation}
\label{eq:vector}A(t) = A_{0}\cos \omega t
\end{equation}
(notice that $m$ here is the free electron mass). In these equations
we have used Bloch's theorem to decouple the Hamiltonian into reduced
components, which we denote by the label $k$. We can transform $H$
into a Floquet Hamiltonian $H_F$ by using as a basis the eigenstates
$|\ell ,n\rangle \equiv $ $|G_\ell +k,n\rangle.$ The Bloch-Floquet
states fully incorporate the symmetries of the original Hamiltonian:
$G$ and $n$ are associate to the discrete space translations of the
lattice vectors and time translations of the vector potential,
respectively. One can then show\cite{jalbert86,brandi88} that the
spectrum of $H$ follows from the diagonalization of the Bloch-Floquet
matrix
\begin{equation}
\label{eq:blochfloquet}\langle \ell ^{\prime },n^{\prime }|H_F|\ell
,n\rangle =\left[ n\hbar \omega +\frac{\hbar ^2(G_\ell+k)^2}{2m}\right]
\delta _{\ell ,\ell ^{\prime }}\delta _{n,n^{\prime }}+J_{n^{\prime
}-n}\negthinspace \Biggl(
\frac{eA_0}{\omega mc}(G_\ell -G_{\ell ^{\prime }})\Biggr)
\sigma_{\ell -\ell ^{\prime }},
\end{equation}
where $J_n$ is the Bessel function of order $n$ and
$\sigma_{\ell}=\sigma$ for $ -3\le \ell \le 3$, and zero otherwise.
The eigenvalues of the Floquet Hamiltonian are quasi-energies but, for
simplicity, will be referred to as energies. The units and conventions
are the same as those defined in Sec.~\ref{sec:2}.

We carry out quantum calculations considering a plane-wave basis set
with a finite number of $G_\ell $ : $ -\ell_{\rm max}\le \ell \le
\ell_{\rm max}$, which leads to a $(2\ell_{\rm max}+1)$--band model in
the absence of the laser field.  We take $\ell_{\rm max}=3$ in most
calculations presented below, which corresponds to a seven-band model.
The choice of a finite number of bands to describe a real crystal is
justified, since highly excited bands (differing from the Fermi level
by an amount larger that the work function of the material) can never
be accessed due to the photoelectric effect. In the absence of the
laser field, the electronic energy levels of (\ref{eq:tothamilt})
consist essentially of the free-electron parabola folded into the
first Brillouin zone, plus gaps opening at the crossing points ($k=0$
and $\pm G_1/2$). Inclusion of the laser field produces a dressing
effect in the bands, which may be described by replica bands
translated by $n\omega_0$ for integer values of $n$. The values
adopted for $\sigma^0$ and $\omega_0$ are such as to strongly mix the
replicas forming the dressed bands. The interaction
$H_{\scriptsize{\mbox{int}}}$ causes an anticrossing whenever two
noninteracting replica bands cross. Our calculations also involve
truncation of the basis set into a finite number of $n$-values:
$|n|\leq n_{\rm max}$. The value of $n_{\rm max}$ is chosen such as to
guarantee the completeness of the replica-bands within the energy
range of interest, and consequently convergence in the calculation of
spectral properties (see discussion below).

\begin{figure}
\setlength{\unitlength}{1mm}
\begin{picture}(100,150)(0,0)
\put(0,20){\epsfxsize=100mm\epsfbox{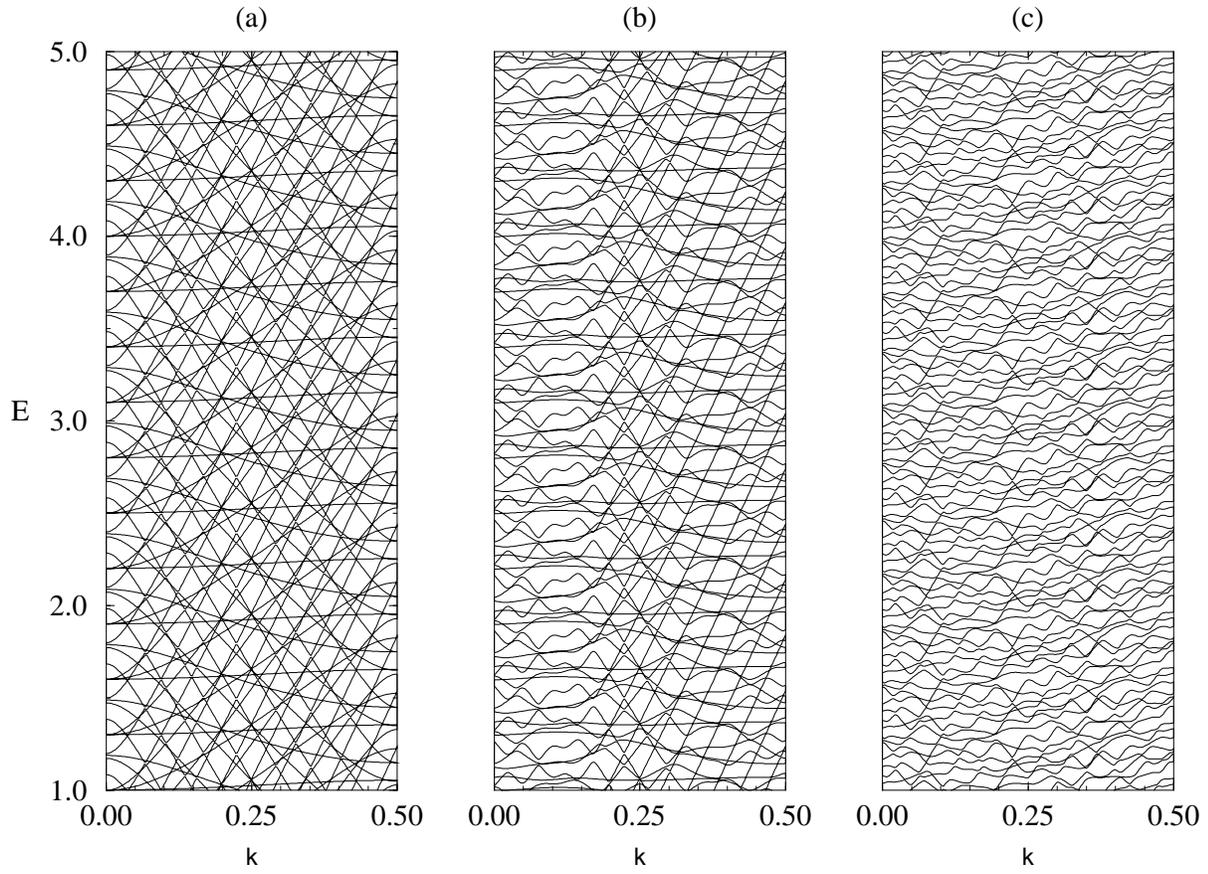}}
\end{picture}
\caption{Energy bands for (a) $x=0$ , (b) $x=1$, and (c) $x=10$ 
calculated with parameters $\ell_{\rm max}= 3$ and $n_{\rm max}= 40$.}
\label{fig4}
\end{figure}

Figure~\ref{fig4} shows the dressed bands spectrum for the energy
range $1<E<5$ at $x=0$, $x=1$, and $x=10$ (as mentioned before, in a
real situation the appropriate range of energy would be defined by the
Fermi energy and the work function of the material). For this energy
range, $n_{\rm max} =40$ is sufficient to achieve saturation in the
replicas. For $x=0$ no level repulsion is observed while for $x=1$
some structure in the energy bands is clearly identified. For $x=10$
the spectrum reveals an intricate level repulsion structure which
resembles those characteristic of quantum chaotic regimes in other
systems \cite{reviewchaos}. All the main features of the classical
stroboscopic map can be inferred from these spectra. For all energies
we can observe that, as $x$ increases, the anticrossing and enhanced
level repulsion among the electronic bands leads to narrower
minibands. From elementary band theory, narrow bands are always
associated to poor transport properties. For example, in a standard
tight-binding scenario, the band width is proportional to the hopping
matrix element between atomic orbitals (equivalent to Wannier
functions in the present plane-waves description) centered at
neighboring lattice sites. Smaller values of this hopping element
indicate a tendency towards localization, which, in the corresponding
classical motion, appears as trajectories confined to a narrower range
of $Q$ values (recall similar discussion in Sec.~\ref{sec:2} with
respect to Fig.~\ref{fig3}).

\begin{figure}
\setlength{\unitlength}{1mm}
\begin{picture}(100,180)(0,0)
\put(30,10){\epsfxsize=60mm\epsfbox{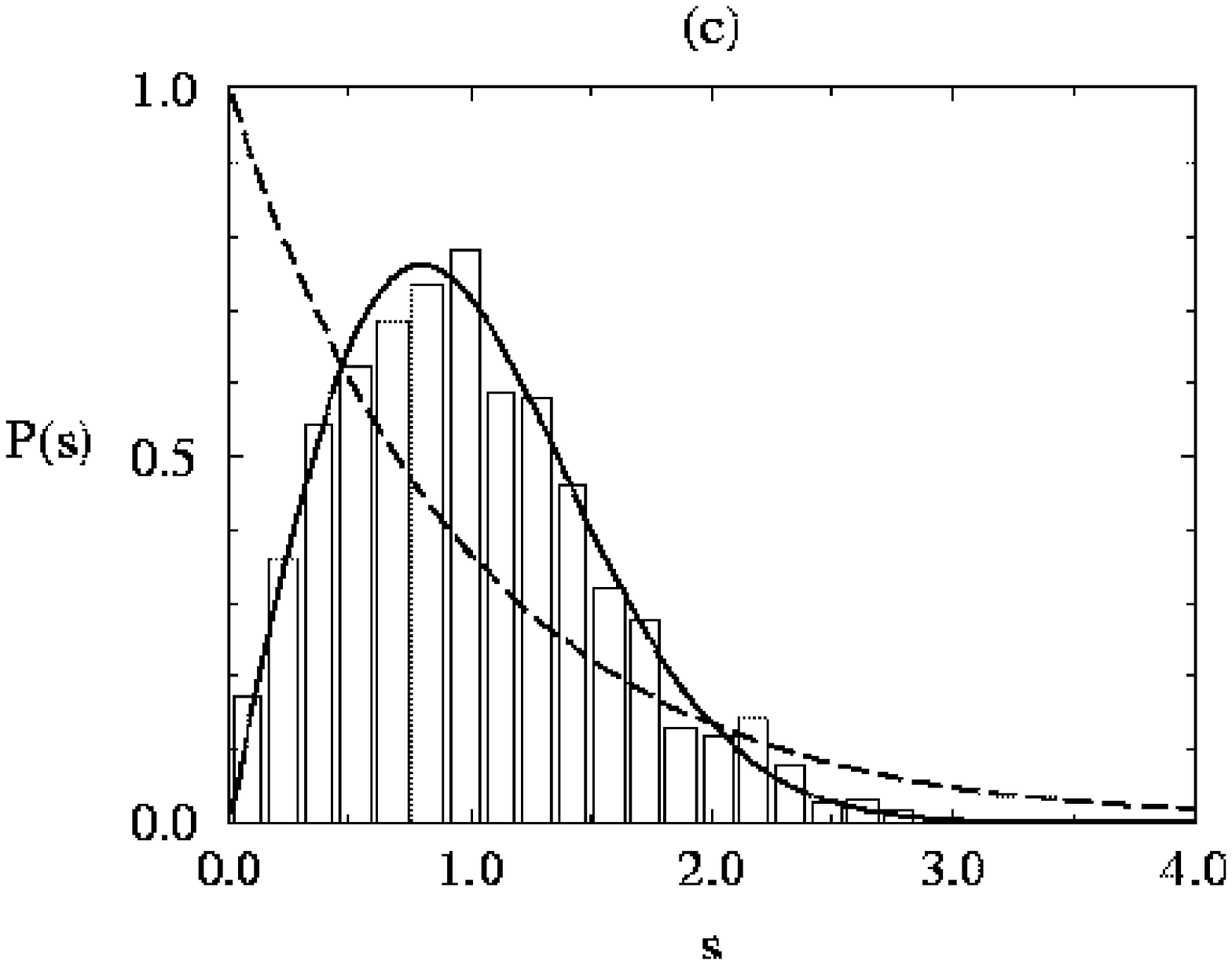}}
\put(30,60){\epsfxsize=60mm\epsfbox{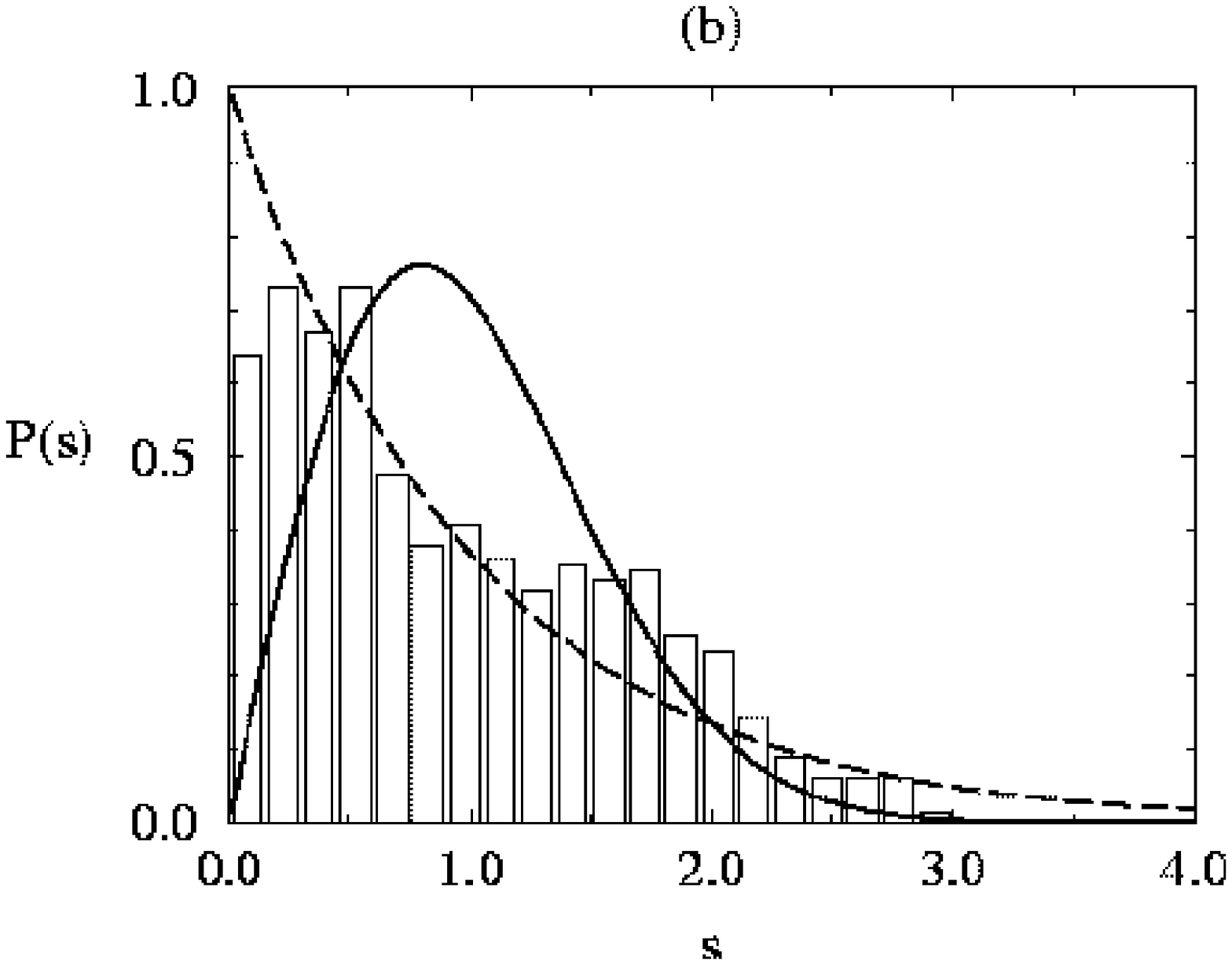}}
\put(30,110){\epsfxsize=60mm\epsfbox{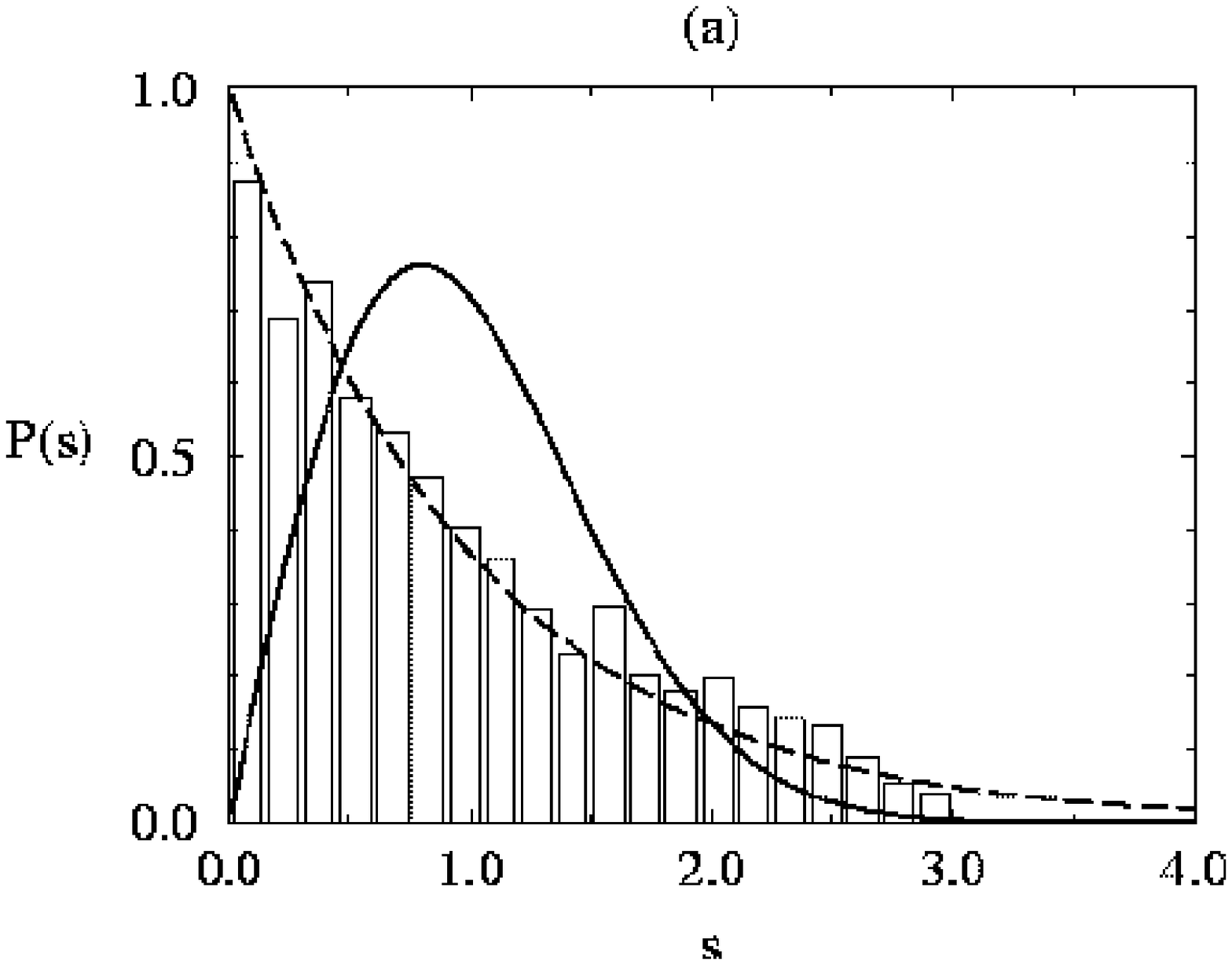}}
\end{picture}
\caption{Nearest neighbors statistics (NNS) for (a) $x=0$, (b) $x=1$, 
and (c) $x=10$ for $\ell_{\rm max}= 3$ and $n_{\rm max}= 40$. The
dashed lines are Poissonian distributions while the solid lines
correspond to GOE distributions of level spacings.}
\label{fig5}
\end{figure}

These trends are confirmed by a quantitative analysis of the spectral
fluctuations. We recall\cite{reviewchaos} that weak and strong
interlevel correlations have been associated with classically regular
and chaotic behaviors, respectively. The latter is usually described
at the quantum level by the Gaussian ensembles of random matrix theory
\cite{mehta91}. In Figure~\ref{fig5} we present the distribution of
nearest-neighbor level spacings (NNS) obtained for increasing values
of the field intensity parameter $x$, with bands confined to the
energy interval 1 to 5. For $x=0$ [Figure~\ref{fig5}(a)] the NNS
distribution is very close to an exponential (the Poisson law),
demonstrating the lack of short-range correlation between levels. The
corresponding classical momentum varies roughly from -3 to 3 for this
energy interval. For $x=1$ [Figure~\ref{fig5}(b)] a crossover regime
appears in which the NNS is neither Poissonian nor GOE-like
\cite{obs}. However, for $x=10$ [Figure~\ref{fig5}(c)] the Wigner
surmise provides a very accurate fit, as expected from the classical
analysis. The curves obtained for the least square deviation
$\Delta_3$ \cite{mehta91} (not shown) lead to a similar interpretation
with respect to large-range correlations.

\begin{figure}
\setlength{\unitlength}{1mm}
\begin{picture}(100,120)(0,0)
\put(30,10){\epsfxsize=60mm\epsfbox{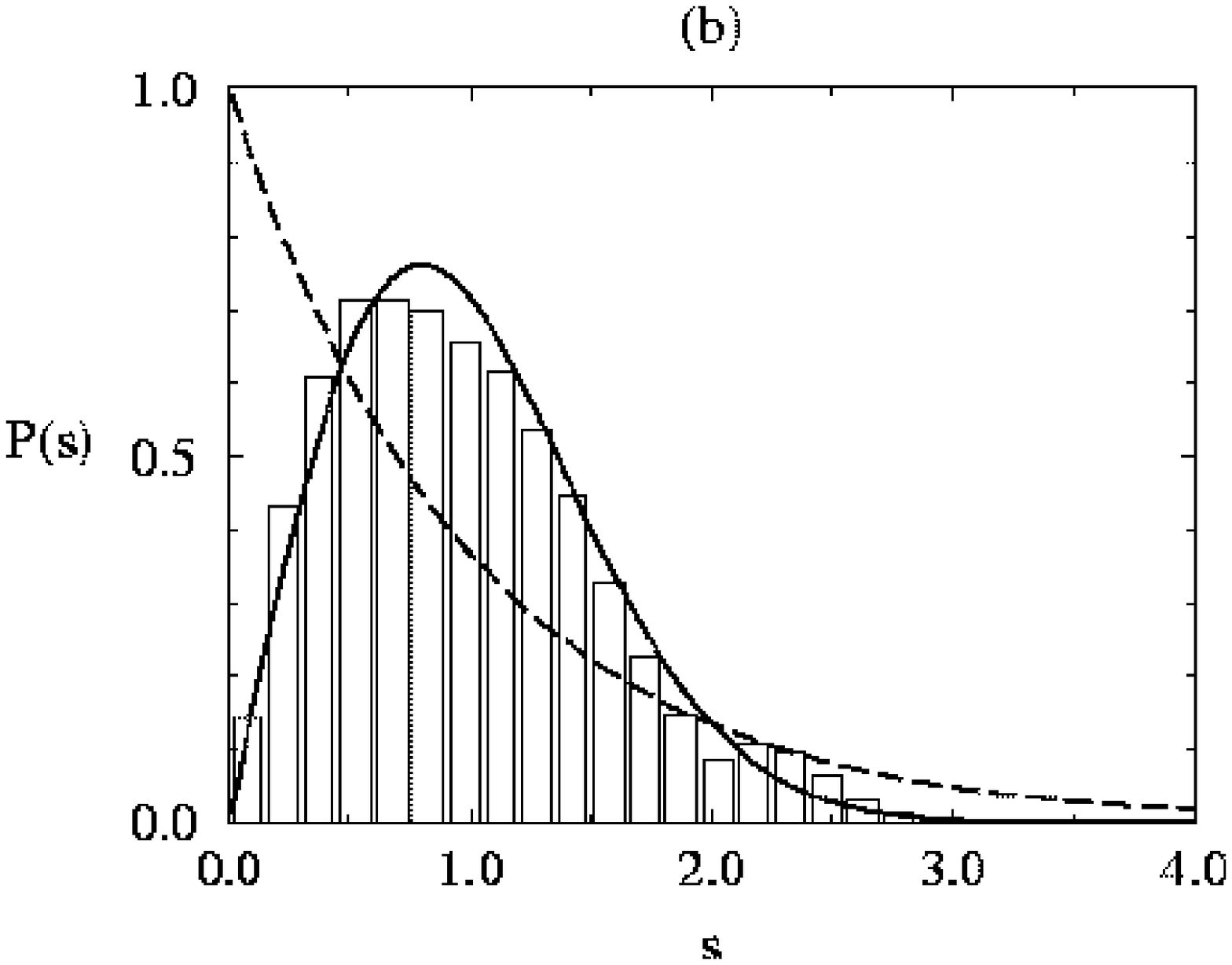}}
\put(30,60){\epsfxsize=60mm\epsfbox{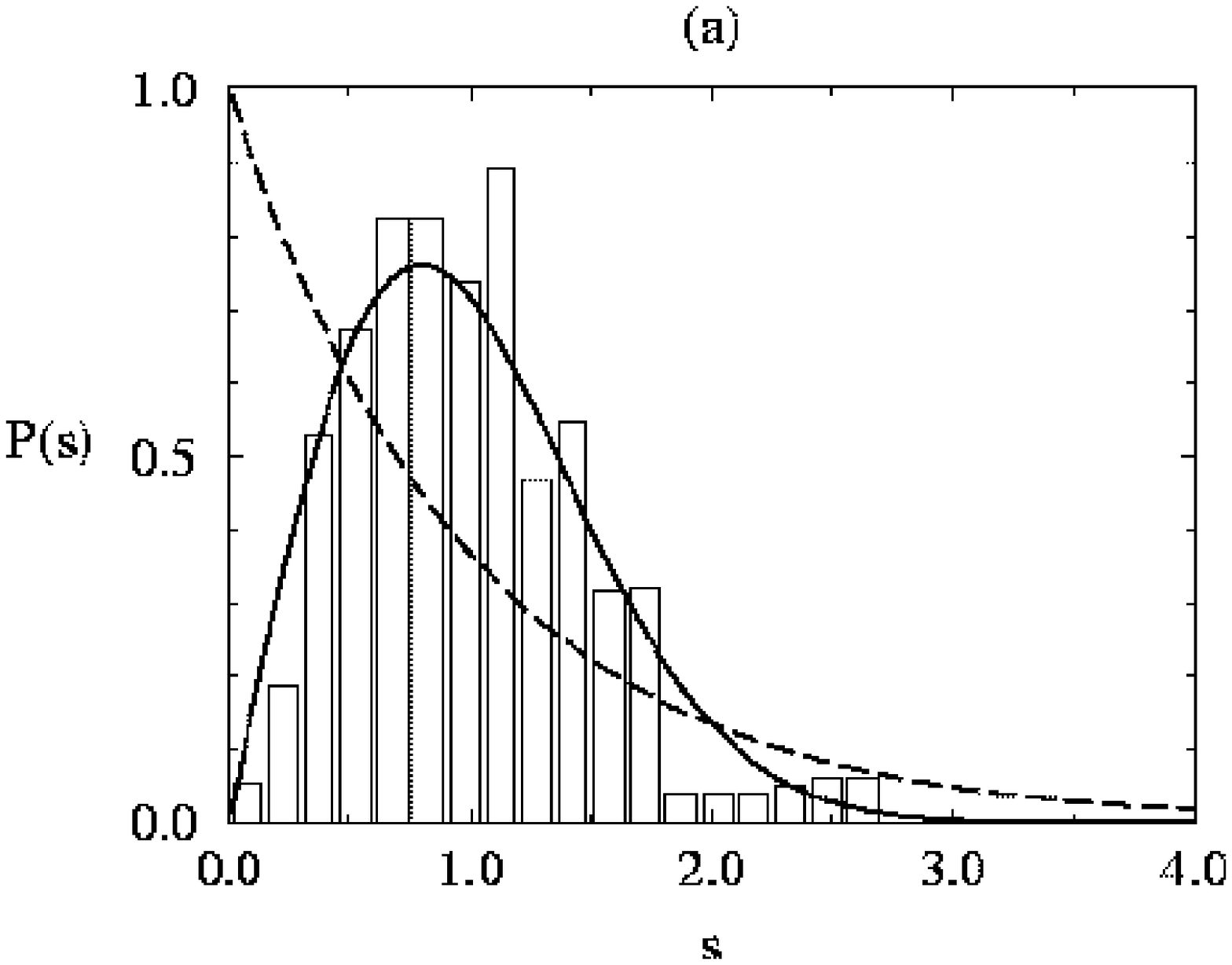}}
\end{picture}
\caption{Dependence of the nearest neighbors statistics (NNS) for
$x=10$ on the number of plane waves $\ell_{\rm max}$. The histograms
correspond to (a) $\ell_{\rm max}= 1$, $n_{\rm max}=40$, and (b)
$\ell_{\rm max}= 5$, $n_{\rm max}=60$, i.e., 3 and 11 electronic bands
models, respectively. These results, together with
Figure~\ref{fig5}(c), illustrate the robustness of our results with
respect to the basis set truncation parameters.}
\label{fig6}
\end{figure}

We now discuss the robustness of our results with respect to the basis
set truncation, i.e., the influence of the cutoff parameters defining
the range of $\ell$ and $n$ values in the calculations presented
above. Figure~\ref{fig6} presents the NNS for $x=10$.
Figure~\ref{fig6}(a) is obtained considering a plane-waves basis
cutoff for $\ell_{\rm max}=1$, while Figure~\ref{fig6}(b) refers to
$\ell_{\rm max}=5$. Note that the results in (a) and (b) are
essentially the same, and also very similar to Figure~\ref{fig5}(c),
illustrating that the plane-wave basis cutoff is not a relevant
parameter. This fact supports that the conclusions drawn from our
model, where only a finite number of bands is incorporated, should be
applicable to real experiments. For a given energy range, $n_{\rm
max}$ must be taken sufficiently large to guarantee
convergence. Since, in the absence of the laser field, the full
spectrum associated to larger values of $\ell_{\rm max}$ covers a
wider range of energies, the adopted value of $n_{\rm max}$ increases
with $\ell_{\rm max}$. Thus $n_{\rm max}=60$ is required if we
consider $\ell_{\rm max}=5$ , while for $\ell_{\rm max}=3$, $n_{\rm
max}=40$ already leads to converged results. The results presented in
Figure~\ref{fig6} show that the choice of the basis set cutoff
parameters must take into account the energy range of physical
interest. Our main conclusions about the system dynamics inferred from
the quantum simulations are well understood from the classical
analysis. Therefore, quantum and classical results are completely
compatible.

%%%%%%%%%%%%%%%%%%%%%%%%%%%%%%%%%%%%%%%%%%%%%%%%%%%%%%%%%%%%%%%%%%%%%%%%%%%%
\section{Experimental aspects}
\label{sec:4}

The scheme proposed here is suitable to model the physics of
high-quality semiconductor vertical superlattices in the presence of
intense monochromatic light radiation \cite{esaki70}. In order to make
the connection we replace the one-dimensional potential by the
superlattice potential. It would not be computationally harder to
introduce a few more reciprocal components to obtain a $V(q)$ in
closer resemblance to that of a superlattice. We found, however, that
increasing the number of components did not change our results
qualitatively. Thus, the only necessary (and crucial) modification is
to replace the free electron mass in Eqs.~(\ref{eq:elechamilt}) and
(\ref{eq:interhamilt}) by the effective mass associated to the
material $m^\ast$. This requires a redefinition of the parameter $x$
as well, namely,
\begin{equation}
x = \left( \frac{\lambda}{d} \right) \left( \frac{m}{m^\ast} \right)
\sqrt{\frac{I}{n_r I_C}}.
\end{equation}
For a specific application, let us assume a CO$_2$ laser with $\lambda
= 10.6\mu$m ($\hbar\omega = 117\mbox{meV}$) and a
$GaAs$-$Al_xGa_{1-x}As$ ($x\approx 0.3$) heterostructure of lattice
constant $d$ = 76\AA. This choice leads to $\hbar\omega =
0.3\epsilon$. The other parameters take the values $m^\ast \approx
0.067m$, $n_r \approx 3.4$, $\epsilon \approx 390\mbox{meV}$, $\sigma
\approx 97\mbox{meV}$, and $I_C \approx 1.2\times 10^{16}$
W/cm$^{2}$. We thus find the relation
\begin{equation}
x^2 \approx \frac{I}{9.4 \times 10^7\ \mbox{W}/\mbox{cm}^{2}}.
\end{equation}
Under these conditions, the laser intensity necessary to produce
strong coupling, namely, $x\approx 10$, is of the order of $10^{10}$
W/cm$^2$. Such intensity is experimentally accessible and does not
cause irreversible damage to the heterostructure
\cite{mysyrowicz86}. Nevertheless, one should have in mind that not
all parameters are fixed by the choice of lattice constant and laser
frequency. The well width can still be chosen independently (implying
in a modification of the ratio $\sigma/\epsilon$). The values chosen
above only represent the strong mixing (chaotic) situation compatible
with our numerical simulations. Of course, decreasing the intensity
would make the system no longer chaotic.

We believe that the transition between regular and chaotic behaviors
can manifest itself in at least two ways. As previously discussed, the
onset of chaos is related to a narrowing of the electronic bands,
causing a decrease in the mobility along the superlattice
direction. As a result, transport measurements should reveal a maximum
of the transverse conductivity when the laser intensity is
small. Another effect of chaos appears in the strong mixing between
energy bands. Optical transitions which were initially forbidden when
no light is shone into the heterostructure should become observable at
high laser field intensities. One could, in principle, irradiate the
superlattice with a pump laser and measure the intensity of a large
set of these new absorption lines using a second (much weaker) probe
laser \cite{mysyrowicz86}. The histogram of oscillator strengths thus
generated should fall into a universal curve predicted by random
matrix theory \cite{taniguchi96} when $x\approx 10$.

%%%%%%%%%%%%%%%%%%%%%%%%%%%%%%%%%%%%%%%%%%%%%%%%%%%%%%%%%%%%%%%%%%%%%%%%%
\section{Conclusions} 
\label{sec:5}

We have studied the interaction of a quantized laser field with an
electron in a 1-D lattice. The classical dynamics of Hamiltonian which
describes the problem was analyzed through the stroboscopic map
technique. We found that a transition from regular to chaotic motion
can be obtained by increasing the laser field intensity. The main
signature of such transition would be a large change in the electron
mobility, leading to a suppression of charge conductance in the
irradiated material. An experimental realization of the model is a
laser interacting with a superlattice. The conditions which must be
fulfilled for the laser and the lattice parameters are compatible with
the experimental possibilities. This implies the use of infrared
lasers and superlattices of periodicity around 100\AA.

%%%%%%%%%%%%%%%%%%%%%%%%%%%%%%%%%%%%%%%%%%%%%%%%%%%%%%%%%%%%%%%%%%%%%%%%%%
\acknowledgments

This work was partially supported by the Brazilian agencies Conselho
Nacional de Desenvolvimento Cient\'\i fico e Tecnol\'ogico (CNPq),
Financiadora de Estudos e Projetos (FINEP), and Funda\c c\~ao
Universit\'aria Jos\'e Bonif\'acio (FUJB-UFRJ).

%%%%%%%%%%%%%%%%%%%%%%%%%%%%%%%%%%%%%%%%%%%%%%%%%%%%%%%%%%%%%%%%%%%%%%%%%%

%%%%%%%%%%%%%%%%%%%%%%%%%%%%%%%%%%%%%%%%%%%%%%%%%%%%%%%%%%%%%%%%%%%%%%%%%%%

\newpage

\newpage 

\newpage

\newpage

\newpage

\end{document}